\documentstyle[12pt]{article}
\newlength{\extraspace}
\newlength{\extraspaces}
\newcommand{\be}{\begin{equation}
\addtolength{\abovedisplayskip}{\extraspaces}
\addtolength{\belowdisplayskip}{\extraspaces}
\addtolength{\abovedisplayshortskip}{\extraspace}
\addtolength{\belowdisplayshortskip}{\extraspace}}
\newcommand{\ee}{\end{equation}}
\newcommand{\ba}{\begin{eqnarray}
\addtolength{\abovedisplayskip}{\extraspaces}
\addtolength{\belowdisplayskip}{\extraspaces}
\addtolength{\abovedisplayshortskip}{\extraspace}
\addtolength{\belowdisplayshortskip}{\extraspace}}
\newcommand{\ea}{\end{eqnarray}}
\newcommand{\nonu}{\nonumber \\[.5mm]}
\newcommand{\A}{&\!\!\!}
\begin{document}
\thispagestyle{empty}
\setlength{\baselineskip}{6mm}
%
%
\vspace*{7mm}
\begin{center}
{\large{\bf Gauge coupling constant, compositeness \\
and \\
supersymmetry}} \\[20mm]
{\sc Kazunari Shima}
\footnote{
\tt e-mail: shima@sit.ac.jp} \ 
and \ 
{\sc Motomu Tsuda}
\footnote{
\tt e-mail: tsuda@sit.ac.jp} 
\\[5mm]
{\it Laboratory of Physics, 
Saitama Institute of Technology \\
Fukaya, Saitama 369-0293, Japan} \\[20mm]
\begin{abstract}
We point out the possibility that the magnitude of the bare gauge coupling constant is determined 
theoretically in a composite model of all particles in SUSY theories.  
We show explicitly in two space-time dimensions that the NL/L SUSY relation 
(i.e., a SUSY compositeness condition for all particles) between $N = 2$ LSUSY QED 
and $N = 2$ NLSUSY determines the magnitude of the bare electromagnetic coupling constant 
(i.e., the fine structure constant) of $N = 2$ LSUSY QED. 
\\[5mm]
\noindent
PACS: 11.30.Pb, 12.60.Jv, 12.60.Rc, 12.10.-g \\[2mm]
\noindent
Keywords: supersymmetry, Nambu-Goldstone fermion, 
nonlinear/linear SUSY relation, composite unified theory 
\end{abstract}
\end{center}

\newpage

\noindent
Supersymmetry (SUSY) \cite{WZ,VA} and the spontaneous SUSY breaking (SSB) are 
crucial notions for the unified description of space-time and matter. 

Based upon the nonlinear (NL) representation \cite{VA} of SUSY and the general relativity (GR) principle 
the nonlinear supersymmmetric general relativity (NLSUSY GR) theory \cite{KS1} is constructed, 
which proposes a new paradigm (called SGM scenario \cite{KS1}-\cite{ST2}) for describing the unity of nature. 
The NLSUSY invariant GR action $L_{\rm NLSUSYGR}(w)$ (in terms of a unified vierbein $w^a{}_\mu$) 
desribes geometrically the basic principle, 
i.e. the ultimate shape of nature is empty unstable space-time with the constant energy density 
(cosmological term) $\Lambda > 0$ and decays to (creates) spontaneously (quantum mechanically) 
ordinary Riemann space-time and spin-${1 \over 2}$ massless Nambu-Goldstone (NG) fermion 
(superon) matter with the potential (cosmological term) $V = \Lambda > 0$ depicted by the SGM action 
$L_{\rm SGM}(e, \psi)$ (in terms of the ordinary vierbein $e^a{}_\mu$ and the NG fermions $\psi^i$). 
We showed explicitly in our previous works 
in asymptotic Riemann-flat ($e^a{}_\mu \rightarrow \delta^a{}_\mu$) space-time 
that the vacuum (true minimum) in SGM scenario is $V = 0$ due to the dynamics of/among the linear (L) SUSY states, 
which is achieved when the NG fermions constitute the LSUSY representation, 
i.e. all local fields of the LSUSY multiplet of the familiar LSUSY gauge theory 
$L_{\rm LSUSY}(A, \lambda, v_a, \dots)$ emerge as the composites of NG fermions 
dictated by the space-time $N$-extended NLSUSY symmetry, 
i.e. LSUSY is realized on the true vacuum as the composite states. 
In the SGM scenario all (observed) particles are assigned uniquely 
into a single irreducible representation of $SO(N)$ ($SO(10)$) super-Poincar\'e (SP) group 
as an on-shell supermultiplet of $N$ LSUSY. 
As mentioned above, they are considered to be realized 
as (massless) eigenstates of $SO(N)$ ($SO(10)$) SP composed of NG fermions (superon) 
which correspond to the coset space coordinates 
for $super-GL(4,R) \over GL(4,R)$ of NLSUSY describing the SSB by itself. 

In order to extract the low energy physical contents of NLSUSY GR and to examine the SGM scenario, 
we investigated the NLSUSY GR model in asymptotic Riemann-flat space-time, 
where the NLSUSY model appears from the cosmological term of NLSUSY GR. 
The NLSUSY model \cite{VA} are recasted (related) to various LSUSY ({\it free}) theories 
with the SSB \cite{IK1}-\cite{ST8} 
and also to {\it interacting} $N = 2$ LSUSY (Yukawa-interaction and QED) theories 
in two dimensional space-time ($d = 2$) for simplicity \cite{ST4-1}-\cite{ST6-2} 
under the adoption of the simplest {\it SUSY invariant constraints} 
and the subsequent {\it SUSY invariant relations} in {\it NL/L SUSY relation}. 
(Note that for $N = 2$ SUSY in SGM scenario $J^P = 1^-$ vector field appears \cite{STT2}. 
Therefore $N = 2$ SUSY is minimally viable case.) 
The SUSY invariant relations, which are obtained {\it systematically} from the SUSY invariant constraints 
\cite{IK1,IK2,UZ} in the superfield formulation \cite{WB}, describe all component fields 
in the LSUSY multiplets as the composites of the NG-fermion superons 
and connect actions between the NLSUSY model and the LSUSY theories with the SSB. 
Through the NL/L SUSY relation, the scale of the SSB in NLSUSY which is related naturally 
to the cosmological constant of NLSUSY GR 
gives a simple explanation of the mysterious (observed) numerical relation 
between the neutrino mass and the four dimensional dark energy density of the universe \cite{ST2,ST5} 
in the vacuum of the $N = 2$ SUSY QED theory \cite{STL}. 

In this letter we study further the NL/L SUSY relation under {\it more general} SUSY invariant constraints 
by considering the $d = 2$, $N = 2$ SUSY QED theory \cite{ST4-2,ST6-2}. 
We find an intimate relation between the bare gauge coupling constant 
and a generalization of SUSY invariant constraints on a general gauge superfield. 
We show explicitly that the magnitude of the bare electromagnetic coupling constant 
is determined in NL/L SUSY relation (i.e. the over-all compositness condition) 
from vacuum expectation values (vevs) (constant terms in SUSY invariant relations) of auxiliary scalar fields, 
which is obtained from the generalized SUSY invariant constraints. 

For the purpose of the brief review of the SUSY invariant constraints in NL/L SUSY relation 
for the $N = 2$ SUSY QED in $d = 2$, 
let us begin with a $d = 2$, $N = 2$ general gauge superfield \cite{DVF,ST7} 
on specific supertranslations of superspace coordinates \cite{IK1,IK2,UZ} 
denoted by $(x'^a, \theta_\alpha'^i)$ ($i = 1, 2$) 
which depend on (Majorana) NG fermions $\psi_\alpha^i(x)$ in the NLSUSY model \cite{VA}, i.e. 
\be
{\cal V}(x', \theta') \equiv \tilde{\cal V}(x, \theta), 
\label{VSFpsi}
\ee
where 
\ba
\A \A 
x'^a = x^a + i \kappa \bar\theta^i \gamma^a \psi^i, 
\nonu
\A \A 
\theta'^i = \theta^i - \kappa \psi^i, 
\ea
with a constant $\kappa$ whose dimension is (mass)$^{-1}$. 
In Eq.(\ref{VSFpsi}), $\tilde{\cal V}(x, \theta)$ may be expanded as 
\ba
\tilde{\cal V}(x, \theta) \A = \A \tilde C(x) + \bar\theta^i \tilde\Lambda^i(x) 
+ {1 \over 2} \bar\theta^i \theta^j \tilde M^{ij}(x) 
- {1 \over 2} \bar\theta^i \theta^i \tilde M^{jj}(x) 
+ {1 \over 4} \epsilon^{ij} \bar\theta^i \gamma_5 \theta^j \tilde\phi(x) 
\nonu
\A \A 
- {i \over 4} \epsilon^{ij} \bar\theta^i \gamma_a \theta^j \tilde v^a(x) 
- {1 \over 2} \bar\theta^i \theta^i \bar\theta^j \tilde\lambda^j(x) 
- {1 \over 8} \bar\theta^i \theta^i \bar\theta^j \theta^j \tilde D(x), 
\label{tVSF}
\ea
where initial component fields in the general gauge superfield ${\cal V}(x, \theta)$, 
i.e. $\varphi_{\cal V}^I(x) = \{ C(x), \Lambda^i(x), M^{ij}(x), \cdots \}$, are denoted 
by $(C, D)$ for two scalar fields, $(\Lambda_\alpha^i, \lambda_\alpha^i)$ for two doublet (Majorana) spinor fields, 
$\phi$ for a pseudo scalar field, $v^a$ for a vector field, 
and $M^{ij} = M^{(ij)}$ $\left(= {1 \over 2}(M^{ij} + M^{ji}) \right)$ 
for three scalar fields ($M^{ii} = \delta^{ij} M^{ij}$), respectively. 
The component fields $\tilde\varphi_{\cal V}^I(x) = \{ \tilde C(x), \tilde\Lambda^i(x), \tilde M^{ij}(x), \cdots \}$ 
in Eq.(\ref{tVSF}) are expressed in terms of the initial $\varphi_{\cal V}^I(x)$ and $\psi^i(x)$ \cite{ST8} 
by using Eq.(\ref{VSFpsi}). 

In Eq.(\ref{tVSF}) (the SUSY invariant) constraints, $\tilde\varphi_{\cal V}^I(x) = {\rm constant}$, 
which eliminate the other degrees of freedom (d.o.f.) than $\varphi_{\cal V}^I(x)$ and $\psi^i(x)$, 
can be imposed, since the superfield (\ref{VSFpsi}) transforms homogeneously \cite{IK1,IK2,UZ} as 
%
$\delta_\zeta \tilde{\cal V}(x, \theta) = \xi^a \partial_a \tilde{\cal V}(x, \theta)$ 
%
with $\xi^a = i \kappa \bar\psi^i \gamma^a \zeta^i$ 
under supertranslations of $(x^a, \theta^i)$ and NLSUSY transformations of $\psi^i(x)$ \cite{VA}, 
\be
\delta_\zeta \psi^i = {1 \over \kappa} \zeta^i 
- i \kappa \bar\zeta^j \gamma^a \psi^j \partial_a \psi^i, 
\label{NLSUSY}
\ee
parametrized by constant (Majorana) spinor parameters $\zeta^i$. 
Then, the simplest (and nontrivial) SUSY invariant constraints are given by \cite{ST8,ST6-1,ST6-2} 
\be
\tilde C = \tilde\Lambda^i = \tilde M^{ij} = \tilde\phi = \tilde v^a = \tilde\lambda^i = 0, 
\ \ \ \tilde D = {\xi \over \kappa} 
\label{SUSYconst0}
\ee
with an arbitrary real paramater $\xi$, 
which give systematically the SUSY invariant relations expressing the component fields $\varphi_{\cal V}^I(x)$ 
as SUSY composites of $\psi^i(x)$ by solving Eq.(\ref{SUSYconst0}); namely, $\varphi_{\cal V}^I = \varphi_{\cal V}^I(\psi;\xi)$. 

As for the component fields of the matter sector of the $N = 2$ SUSY QED, 
we perform the similar arguments for two $d = 2$, $N = 2$ scalar superfields on $(x'^a, \theta'^i)$ defined by 
\be
\Phi^i(x', \theta') \equiv \tilde \Phi^i(x, \theta), 
\label{SSFpsi}
\ee
and 
\be
\tilde \Phi^i(x, \theta) 
= \tilde B^i(x) + \bar\theta^i \tilde \chi(x) - \epsilon^{ij} \bar\theta^j \tilde \nu(x) 
- {1 \over 2} \bar\theta^j \theta^j \tilde F^i(x) + \bar\theta^i \theta^j \tilde F^j(x) + \cdots, 
\label{tSSF}
\ee
where initial component fields in the scalar superfields $\Phi^i(x, \theta)$, 
i.e. $\varphi_\Phi^I(x) =$ \{$B^i(x)$, $\chi(x)$, $\nu(x)$, $F^i(x)$\}, 
are denoted by $B^i$ for doublet scalar fields, 
$(\chi_\alpha, \nu_\alpha)$ for two (Majorana) spinor fields and $F^i$ for auxiliary scalar fields. 
The superfields (\ref{tSSF}) also satisfy the same supertransformation property 
$\delta_\zeta \tilde \Phi^i(x, \theta) = \xi^a \partial_a \tilde \Phi^i(x, \theta)$ 
and constraints, $\tilde\varphi_\Phi^I(x) = {\rm constant}$, 
are invariant (conserved quantities) under the supertransformations. 
The simplest SUSY invariant constraints are \cite{ST6-2} 
\be
\tilde B^i = \tilde\chi = \tilde\nu = 0, \ \ \ \tilde F^i = {\xi^i \over \kappa} 
\label{SUSYconst-SSF}
\ee
with arbitrary real paramaters $\xi^i$, which give SUSY invariant relations for the all component fields 
$\varphi_\Phi^I(x)$; namely, $\varphi_\Phi^I = \varphi_\Phi^I(\psi;\xi^i)$. 

Here we consider the simplest generalization of the SUSY invariant constraints (\ref{SUSYconst0}) 
by taking the following particular (non-zero) constant value for the auxiliary scalar field $\tilde C(x)$ as 
\be
\tilde C = \xi_c, 
\label{SUSYconst-C}
\ee
where $\xi_c$ is a constant whose dimension is (mass)$^{-1}$ in $d = 2$ (a dimensionless constant in $d = 4$). 
This generalization of the SUSY invariant constraints (\ref{SUSYconst0}) means that 
the SUSY invariant relation, $C = C(\psi;\xi)$, is shifted by $\xi_c$ as, 
\be
C = \xi_c - {1 \over 8} \xi \kappa^3 \bar\psi^i \psi^i \bar\psi^j \psi^j, 
\label{SUSYrelation-C}
\ee
which is obtained by solving Eq.(\ref{SUSYconst-C}) together with the other SUSY invariant constraints 
in Eq.(\ref{SUSYconst0}), 
\be
\tilde\Lambda^i = \tilde M^{ij} = \tilde\phi = \tilde v^a = \tilde\lambda^i = 0, 
\ \ \ \tilde D = {\xi \over \kappa}. 
\label{SUSYconst}
\ee
Note that the generalized constraint (\ref{SUSYconst-C}) does not alter the other SUSY invariant relations 
$\varphi_{\cal V}^I = \varphi_{\cal V}^I(\psi;\xi)$ than Eq.(\ref{SUSYrelation-C}). 

Let us show below the relation between a $N = 2$ NLSUSY action and a general $N = 2$ SUSY QED action 
under the adoption of the generalized SUSY invariant constraint (\ref{SUSYconst-C}) 
in addition to Eqs.(\ref{SUSYconst-SSF}) and (\ref{SUSYconst}). 
The $N = 2$ NLSUSY action \cite{VA} which is invariant under the NLSUSY transformations (\ref{NLSUSY}) is given by 
\be
L_{N = 2{\rm NLSUSY}} = - {1 \over {2 \kappa^2}} \ \vert w \vert, 
\label{NLSUSYaction}
\ee
where $\vert w \vert$ is the determinant describing the dynamics of $\psi^i(x)$, 
i.e. in $d = 2$, 
\be
\vert w \vert = \det(w^a{}_b) = \det(\delta^a_b + t^a{}_b) 
= 1 + t^a{}_a + {1 \over 2!}(t^a{}_a t^b{}_b - t^a{}_b t^b{}_a) 
\label{det-w}
\ee
with $t^a{}_b = - i \kappa^2 \bar\psi^i \gamma^a \partial_b \psi^i$. 
In the NLSUSY action (\ref{NLSUSYaction}) the second term, 
$-{1 \over {2 \kappa^2}} t^a{}_a = {i \over 2} \bar\psi^i \!\!\not\!\partial \psi^i$, 
is the kinetic term for $\psi^i$. 

On the other hand, the general $N = 2$ SUSY QED action \cite{ST6-2} in $d = 2$ is constructed 
from the $N = 2$ general gauge and $N = 2$ scalar superfields, 
${\cal V}(x, \theta)$ and $\Phi^i(x, \theta)$, as 
\be
L^{\rm gen.}_{N = 2{\rm SUSYQED}} 
= L_{{\cal V}{\rm kin}} + L_{{\cal V}{\rm FI}} 
+ L_{\Phi{\rm kin}} + L_e 
\label{SQEDaction}
\ee
with 
\ba
L_{{\cal V}{\rm kin}} 
\A = \A {1 \over 32} \left\{ \int d^2 \theta^i 
\ (\overline{D^i {\cal W}^{jk}} D^i {\cal W}^{jk} 
+ \overline{D^i {\cal W}_5^{jk}} D^i {\cal W}_5^{jk}) \right\}_{\theta^i = 0}, 
\label{Vkin}
\\
L_{{\cal V}{\rm FI}} 
\A = \A {\xi \over {2 \kappa}} \int d^4 \theta^i \ {\cal V}, 
\label{VFI}
\\
L_{\Phi{\rm kin}} + L_e 
\A = \A - {1 \over 16} \int d^4 \theta^i \ e^{-4e{\cal V}} (\Phi^j)^2, 
\label{gauge}
\ea
where we denote the kinetic terms for the vector supermultiplet by $L_{{\cal V}{\rm kin}}$ 
with ${\cal W}^{ij} = \bar D^i D^j {\cal V}$, ${\cal W}_5^{ij} = \bar D^i \gamma_5 D^j {\cal V}$, 
$D^i = {\partial \over \partial\bar\theta^i} - i \!\!\not\!\!\partial \theta^i$, 
the Fayet-Iliopoulos (FI) $D$ term indicating the SSB by $L_{{\cal V}{\rm FI}}$, 
and the kinetic terms for the matter scalar supermultiplet 
and the gauge interaction terms by $L_{\Phi{\rm kin}} + L_e$, respectively. 
The bare gauge coupling constant is defined in the gauge action (\ref{gauge}) by $e$ 
whose dimension is $({\rm mass})^1$ in $d = 2$, 

Substituting all the generalized SUSY invariant relations by Eq.(\ref{SUSYrelation-C}), 
$\varphi_{\cal V}^I = \varphi_{\cal V}^I(\psi;\xi,\xi_c)$ and $\varphi_\Phi^I = \varphi_\Phi^I(\psi;\xi^i)$, 
into the $N = 2$ SUSY QED action (\ref{SQEDaction}), 
we find the (general) NL/L SUSY relation for the general off-shell supermultiplet: 
\be
L^{\rm gen.}_{N = 2{\rm SUSYQED}} =  f(\xi, \xi^i, \xi_c, e) \ L_{N = 2{\rm NLSUSY}}, 
\label{NLSUSY-genQED}
\ee
where the normalization factor $f(\xi, \xi^i, \xi_c, e)$ is given by 
\be
f(\xi, \xi^i, \xi_c, e)= \xi^2 - (\xi^i)^2 e^{-4 e \xi_c}.  
\label{f-xi}
\ee
In the factor (\ref{f-xi}) $\xi_c$, $\xi$ and $\xi^i$ are the real parameters 
giving the magnitude of the vevs (constant terms in SUSY invariant relations) for the auxiliary fields, 
($C$, $D$, $F^i$) in ${\cal V}$ and $\Phi^i$, respectively. 
The NL/L SUSY relation (\ref{NLSUSY-genQED}) for $N = 2$ is obtained straightforwardly (and systematically) 
by changing the integration variables \cite{ST8} in the general $N = 2$ SUSY QED action (\ref{SQEDaction}) 
from $(x, \theta^i)$ to $(x', \theta'^i)$ 
under the SUSY invariant constraints (\ref{SUSYconst-SSF}), (\ref{SUSYconst-C}) and (\ref{SUSYconst}). 

Here we remind ourselves of the NLSUSY GR model in SGM scenario. 
In the SGM action $L_{\rm SGM}(e,\psi)$ 
the relative scale of the kinetic terms for the NG fermions $\psi^i$ to the Einstein GR theory is fixed as 
$-{1 \over {2 \kappa^2}} t^a{}_a = {i \over 2} \bar\psi^i \!\!\not\!\!\partial \psi^i$; 
namely, the SGM action can be written (in $d = 4$) as \cite{KS1}-\cite{ST2} 
\be
L_{\rm SGM}(e,\psi) = {c^4 \over {16 \pi G}} e ( R - \vert w \vert \Lambda + \cdots ), 
\label{SGM}
\ee
and then the dimensional constant $\kappa$ in the determinant (\ref{det-w}) 
is fixed to $\kappa^{-2} = {{c^4 \Lambda} \over {8 \pi G}}$. 
In order to realize the correct kinetic terms of $\psi^i$ 
(and the cosmological term) in $L_{\rm SGM}(e,\psi)$ from the NL/L SUSY relation (\ref{NLSUSY-genQED}), 
i.e. since it is essential for the SGM scenario to attribute exactly a LSUSY action 
with the SSB (for the low energy physics) to the NLSUSY action corresponding to 
the SGM action for flat space-time, we put 
\be
f(\xi, \xi^{i}, \xi_c, e) = 1,  
\label{f-xi-1}
\ee
which establishes the NL/L SUSY relation (in SGM scenario) for $N = 2$ SUSY. 
The fact that the condition (\ref{f-xi-1}) reproduces correctly the SGM action (\ref{SGM}) for flat space-time 
means that a configration of the vevs for the auxiliary fields satisfying $f(\xi, \xi^{i}, \xi_c, e) = 1$ 
exists based on the fundamental theory, which is natural for the SGM composite scenario. 
This is different from the situation in string theory where a compactification of extra dimensions 
contributes essentially to the magnitude of the coupling constant. 

Remarkably the condition (\ref{f-xi-1}) gives the gauge coupling constant $e$ 
in terms of $\xi$, $\xi^i$ and $\xi_c$ as 
\be
e = {1 \over 4\xi_c} \ln X, 
\ \ \ X = {(\xi^i)^2 \over {\xi^2 - 1}}. 
\label{gcoupling}
\ee

Now we discuss the NL/L SUSY relation in the $d = 2$, $N = 2$ LSUSY QED theory 
for a {\it minimal} off-shell vector supermultiplet \cite{ST4-2,ST6-2} 
based on the (generalized) NL/L SUSY relation (\ref{NLSUSY-genQED}). 
We ask that a local $U(1)$ gauge invariant 
and $N = 2$ LSUSY QED action for the minimal off-shell vector supermultiplet, 
\be
L^0_{N = 2{\rm LSUSYQED}} 
= L^0_{{\cal V}{\rm kin}} + L^0_{{\cal V}{\rm FI}} + L^0_{\Phi{\rm kin}} + L^0_e, 
\label{minSQEDaction}
\ee
where 
\ba
L^0_{{\cal V}{\rm kin}} 
\A = \A - {1 \over 4} (F_{0ab})^2 
+ {i \over 2} \bar\lambda_0^i \!\!\not\!\partial \lambda_0^i 
+ {1 \over 2} (\partial_a A_0)^2 
+ {1 \over 2} (\partial_a \phi_0)^2 
+ {1 \over 2} D_0^2, 
\label{minSQED1}
\\
L^0_{{\cal V}{\rm FI}} \A = \A - {\xi \over \kappa} D_0, 
\label{minSQED2}
\\
L^0_{\Phi{\rm kin}} 
\A = \A {i \over 2} \bar\chi \!\!\not\!\partial \chi 
+ {1 \over 2} (\partial_a B^i)^2 
+ {i \over 2} \bar\nu \!\!\not\!\partial \nu 
+ {1 \over 2} (F^i)^2, 
\\
L^0_e 
\A = \A e \ \bigg\{ i v_{0a} \bar\chi \gamma^a \nu 
- \epsilon^{ij} v_0^a B^i \partial_a B^j 
+ \bar\lambda_0^i \chi B^i 
+ \epsilon^{ij} \bar\lambda_0^i \nu B^j 
- {1 \over 2} D_0 (B^i)^2 
\nonu
\A \A 
+ {1 \over 2} A_0 (\bar\chi \chi + \bar\nu \nu) 
- \phi_0 \bar\chi \gamma_5 \nu \bigg\} 
+ {1 \over 2} e^2 (v_{0a}{}^2 - A_0^2 - \phi_0^2) (B^i)^2, 
\label{minSQED4}
\ea
should be reproduced from Eq.(\ref{NLSUSY-genQED}). 

In Eqs.(\ref{minSQED1}), (\ref{minSQED2}) and (\ref{minSQED4}), 
gauge invariant quantities \cite{ST6-1,WB} corresponding to the d.o.f. for the minimal off-shell vector supermultiplet 
are denoted by 
\be
\{ A_0, \phi_0, F_{0ab}, \lambda_0^i, D_0 \} 
\equiv \{ M^{ii}, \phi, F_{ab}, \lambda^i + i \!\!\not\!\partial \Lambda^i, D + \Box C \}, 
\label{gauge-inv}
\ee
with $F_{0ab} = \partial_a v_{0b} - \partial_b v_{0a}$ and $F_{ab} = \partial_a v_b - \partial_b v_a$, 
which are invariant ($v_{0a} = v_a$ transforms as an Abelian gauge field) 
under a SUSY generalized gauge transformation in the LSUSY theory, 
$\delta_g {\cal V} =\Lambda^1 + \alpha \Lambda^2$ with the real constant $\alpha$ 
and the generalized  $N = 2$ scalar superfield gauge parameter $\Lambda^i$. 
The fields $\{ A_0, \phi_0, v_{0a}, \lambda_0^i, D_0 \}(\psi;\xi)$ constitute 
the minimal off-shell vector supermultiplet of $N = 2$ SUSY QED 
in NL/L SUSY relation as shown in Refs.\cite{ST6-1,ST6-2}.  

We can see easily that the general $N = 2$ SUSY QED action (\ref{SQEDaction}) related (equivalent) to 
the $N = 2$ NLSUSY action (\ref{NLSUSYaction}) reduces to the minimal $N = 2$ LSUSY QED action (\ref{minSQEDaction}) 
with the arbitrary $\xi_c$ under the bare gauge coupling constant (\ref{gcoupling}). 
Now by adopting the Wess-Zumino gauge for LSUSY we can gauge away all auxiliary fields except $D_0$ in the vector supermultiplet 
and by rescaling the whole scalar supermultiplet $\Phi^i$ in the gauge action (\ref{gauge}) by the constant numerical factor 
$e^{-2e\xi_c}$ we obtain ordinary $N = 2$ SUSY QED action for the {\it minimal} off-shell vector supermultiplet \cite{ST6-1,ST6-2} 
with Eq.(\ref{gcoupling}), which completes the NL/L SUSY relation between the $N = 2$ NLSUSY model and the $N = 2$ LSUSY QED theory 
for the minimal off-shell vector supermultiplet: 
\be
L_{N = 2{\rm NLSUSY}} = L^{\rm gen.}_{N = 2{\rm SUSYQED}} 
= L^0_{N = 2{\rm LSUSYQED}} + [{\rm tot.\ der.\ terms}]. 
\label{NLSUSY-LSQED}
\ee
Interestingly $e$ defined by the action (\ref{gauge}) depends upon the vevs of the auxiliary fields, i.e. 
the vacuum structures. 
(Note that the bare $e$ is a free independent parameter, provided  $\xi_c=0$ as in the case of adopting 
the Wess-Zumino gauge throughout the arguments.) 
%

Now we summarize the results as follows. 
We have considered the simplest generalization of the SUSY invariant constraints 
as Eq.(\ref{SUSYconst-C}) and found the (generalized) NL/L SUSY relation (\ref{NLSUSY-genQED}) 
between the $N = 2$ NLSUSY action and the general $N = 2$ SUSY QED action in $d = 2$, 
which produces the factor (\ref{f-xi}) giving the interesting relation 
between the gauge coupling constant and the vevs (constant terms) of the auxiliary fields 
in Eqs.(\ref{f-xi-1}) and (\ref{gcoupling}).  
Our study may indicate that the structure of the auxiliary fields for the general (gauge) superfield 
plays a crucial role in SUSY theory by determining not only the true vacuum through the SSB due to $D$ term 
but also the magnitude of the (bare) gauge coupling constant through NL/L SUSY relations 
(i.e. the SUSY compositeness condition for all particles and the auxiliary fields as well), 
which is favorable to the SGM scenario for unity of nature. 
The strength of the bare gauge coupling constant may ought to be predicted, 
provided the fundamental theory is that of everything. 
The similar arguments in $d = 4$ for more general SUSY invariant constraints 
and for the large $N$ SUSY, especially $N = 4, 5$ are interesting and crucial.

\newpage

%
\newcommand{\NP}[1]{{\it Nucl.\ Phys.\ }{\bf #1}}
\newcommand{\PL}[1]{{\it Phys.\ Lett.\ }{\bf #1}}
\newcommand{\CMP}[1]{{\it Commun.\ Math.\ Phys.\ }{\bf #1}}
\newcommand{\MPL}[1]{{\it Mod.\ Phys.\ Lett.\ }{\bf #1}}
\newcommand{\IJMP}[1]{{\it Int.\ J. Mod.\ Phys.\ }{\bf #1}}
\newcommand{\PR}[1]{{\it Phys.\ Rev.\ }{\bf #1}}
\newcommand{\PRL}[1]{{\it Phys.\ Rev.\ Lett.\ }{\bf #1}}
\newcommand{\PTP}[1]{{\it Prog.\ Theor.\ Phys.\ }{\bf #1}}
\newcommand{\PTPS}[1]{{\it Prog.\ Theor.\ Phys.\ Suppl.\ }{\bf #1}}
\newcommand{\AP}[1]{{\it Ann.\ Phys.\ }{\bf #1}}

\end{document}